\documentstyle{mn}
\newcommand{\uchiir}{{UCH{\sc ii}R}}
\newcommand{\hii}{{H{\sc\,ii}}}
\newcommand{\cm}{\,{\rm cm}\,}

\newcommand{\second}{\,{\rm s}\,}
\newcommand{\Rf}{{R_{\rm f}}}
\title[Clumpy ultracompact H{\small\it\,II} regions II]
{Clumpy ultracompact H{\Large\bf\,II} regions -- II.\\
Cores, spheres and shells from subsonic flows}
\author[M.P. Redman, R.J.R. Williams and  J.E. Dyson]
{M.P. Redman$^1$, R.J.R. Williams$^1$ and J.E. Dyson$^{1,2}$\\
$^1$Department~of~Physics~and~Astronomy,
University~of~Manchester, Oxford~Road, Manchester M13~9PL\\
$^2$Department~of~Physics~and~Astronomy,
University~of~Leeds, Leeds LS2 9JT}

\date{Accepted 1995 November 27, Received 1995 November 15; in original form 1995 September 27}
\pagerange{\pageref{firstpage}--\pageref{lastpage}}

\begin{document}
\input{epsf}
\label{firstpage}
\maketitle

\begin{abstract}
We have modelled ultracompact H\,{\sc ii} regions (\uchiir) in terms of
steady subsonic ionized flows in a clumpy medium.  Mass loss from
neutral clumps allows the regions to be long-lived.  We examine the
form of global flows for different dependences of the volume mass
injection rate, $\dot{q}$, on radius and Mach number, and describe the
solutions in detail.  We find that three observed \uchiir\
morphologies are reproduced with these models.  Mach number
independent flows that include a radial variation can give centre-
brightened core--halo morphologies.  Mach number dependent flows
reproduce naturally the uniform \uchiir\ morphology.  In a hybrid
model, including subsonic and supersonic flows, we allow a supersonic
wind to shock in the ionized region.  The ionized subsonic gas has a
high density and so dominates the emission.  The shell produced has a
velocity structure very different from that of fully supersonic
models. Several morphologies of spherical \uchiir\ can be understood
in terms of these various models; however, kinematic data are crucial
as a discriminant between them.
\end{abstract}

\begin{keywords}
hydrodynamics -- shock waves -- stars: mass-loss -- ISM: structure -- 
H{\sc ii} regions -- radio lines: ISM
\end{keywords}
\section{introduction}
The importance of massive stars in the Galaxy is hard to
overstate.  Over their relatively brief lifetimes they inject large
amounts of energy and momentum into the interstellar medium through
their UV radiation fields and high-velocity winds.  Eventually they
explode as supernovae, adding heavy elements, and driving shockwaves
into their local environment.  To understand the global properties of
our Galaxy a detailed knowledge of the evolution of massive stars is
essential.  Because of their powerful winds and radiation fields, OB
stars soon disrupt their natal environment, making it difficult to
observe the conditions in which they formed.  However, ultracompact
H\,{\sc ii} regions (\uchiir) are promising objects of study as they
contain relatively young OB stars that have not yet dispersed the
original cocoon of gas from which they formed.

\uchiir\ are embedded deep within molecular clouds and are obscured by dust so that they can only be observed in the radio and far-infrared (FIR) wavebands.
At these wavelengths, they are amongst the brightest compact sources
in the Galaxy.  The typical characteristics of \uchiir, as reviewed by
Churchwell~\shortcite{churchwell.rev90} and Kurtz, Churchwell \& Wood
\shortcite[hereafter KCW]{kurtz.et.al94}, are that \uchiir\ have
small diameters $\la10^{17}\cm$, are dense ($\langle n_{\rm
e}^2\rangle^{1/2}\ga10^5 \cm^{-3}$) and have high emission measures
($\langle n_{\rm e}^2\rangle L\ga10^7 \cm^{-6}{\rm\,pc}$). $L$ is the
distance along the line of sight in parsecs and $n_{\rm e}$ is the
number density of electrons.

Wood \& Churchwell \shortcite{wood&churchwell89a} proposed that
\uchiir\ morphologies could be described by a few types: cometary
($\sim 20$ per cent), core--halo ($\sim 16$ per cent), shell ($\sim 4$
per cent), irregular or multiply peaked ($\sim 17$ per cent) and
spherical/unresolved ($\sim 43$ per cent).  \uchiir\ vary greatly in
appearance, and the assignment of a particular morphological type may,
in many cases, be rather subjective.  In particular, surveys take
snapshots of individual regions at only one or two different spatial
scales, and interferometers such as the VLA can miss large-scale
structures.  Recent long-duration, multiple VLA configuration
observations of Sgr B2 \cite{gaume.et.al95,depree.et.al95b}
dramatically illustrate the complicated interactions between stars and
their local environment.

Most attention has been directed towards the cometary \uchiir\@.  The
steady-state bow shock model
\cite{vanburen.et.al90,maclow.et.al91,vanburen&maclow92} has been
questioned recently.  For example, two well-known cometary \uchiir\
(including the prototypical cometary \uchiir, G34.3~+~0.2C) have been
shown to have a tri-limbed tail structure with a large velocity
gradient perpendicular to the head--tail axis
\cite{gaume.et.al94,gaume.et.al95}.  These observations are
incompatible with the bow shock model.  Moreover, the bow shock model
cannot reproduce the many arc-like \uchiir\ seen in the new images.
We model such arc-like \uchiir\ in a forthcoming paper \cite[Paper
III]{williams.et.al96}.

Dyson \shortcite{dyson94} suggested that the clumpy nature of
molecular clouds could account for the relatively long inferred
lifetimes of \uchiir.  The clumps act as localized sources of mass
which is added slowly to the flow by photoionization and/or
hydrodynamic ablation.  This continuous mass injection leads to a
recombination front bounded \uchiir\ that does not expand quickly, and
so does not lead to the lifetime problems encountered if \uchiir\ are
modelled as `classical' H\,{\sc ii} regions.  In a previous paper
\cite[hereafter Paper I]{dyson.et.al95}, we examined one of the
simple models outlined by Dyson \shortcite{dyson94}.  We calculated
line profiles and emission measures for a supersonic wind driven
\uchiir.  In this model, the ionized flow remains supersonic through
to a recombination front.  This type of solution reproduces the shell
morphology of some \uchiir\ and predicts highly characteristic broad,
double-peaked line profiles.  In the present paper, we further develop
this clumpy environment model and show that, taken with Paper I, there
is reason to believe that these models can explain a significant
fraction of \uchiir\ morphologies.

In Section 2, we describe a model for \uchiir\ in which the dynamical
effects of a central stellar wind are negligible, so that the flow
generated by mass injection is subsonic throughout the ionized region.
This model is mainly appropriate for early B and perhaps late O
main-sequence stars which have less wind momentum than the massive OB
supergiants assumed to be the exciting stars in Paper I\@.  Mass loss
from clumps is driven by photoionization and/or hydrodynamic ablation.
The former gives a mass injection rate that is independent of flow
speed whereas, in the latter case, loading is suppressed at low Mach
numbers \cite{hartquist.et.al86}.  We examine these two possibilities
separately and allow for possible radial variations in the mass
loading rate by including a power-law dependence of the mass injection
rate.  Density and velocity plots, and line profiles and emission
measures are produced for several cases.  We highlight the
similarities and differences between the flows produced by the two
different mass loading laws and show that the uniform spherical
\uchiir\ morphology is naturally reproduced along with the
centre-brightened core--halo type.

In Section 3 a hybrid model is described.  A supersonic central
stellar wind source is mass loaded to such an extent that it shocks
and becomes subsonic before reaching the recombination front. We
produce an example of a line profile and emission measure plot for the
case of a shock occurring close to the recombination front.  We find
that a shell morphology is produced if the emission from the
high-speed, low-density supersonic gas is negligible.  The velocity
structure in this case differs completely from that of the models of
Paper I which have a similar overall morphology, thus providing an
observational test between the two.  We note that it is likely that
shell morphology \uchiir\ resulting from a fully supersonic mass
loaded wind (Paper I) will in reality appear more clumpy than those
that result from the partly subsonic structures described here.

\section{Subsonic and transonic \uchiir\ }
We assume that the flow is subsonic throughout the ionized region.  If
the mass loading is dominated by photoionization, the rate is Mach
number independent, while, if hydrodynamical ablation is the dominant
mechanism, phenomenological arguments suggest that the volume mass
loading rate varies as $M^{4/3}$\cite{hartquist.et.al86}.  We discuss
these two cases separately, but one should note that there may be more
than one mode of mass injection at work in any given flow.  The mass
injection rate is also given a dependence on radial distance in both
cases.  We ignore gravitational effects since $2G M_{*} /c^2 \la
0.04R$ for an $M_{*}\simeq 20\,{\rm M}_{\odot}$ star within an ionized region
of characteristic radius $R=10^{17}\cm$ and an isothermal sound speed
$c\simeq 10~{\rm km}\second^{-1}$.  We also assume that the flow is
dust free, deferring a consideration of the dynamical effects of
radiation pressure on a dust--gas mixture to a later paper (Williams,
Dyson \& Redman, in preparation).  Finally, we assume that the
dominant source of mass injection is from the clouds.  We take a
volume mass injection rate
\begin{equation}
\dot q=\dot{q_0}M^{\beta}(r/\Rf)^{\alpha},
\end{equation}
where $r$ is the radial coordinate, $\Rf$ is the distance of the
recombination front from the star and $\alpha$ and $\beta$ are
constants.

The continuity and momentum equations for isothermal flow are
respectively
\begin{equation}
\label{new.mass}
\frac{{\rm d}}{{\rm d}r}(r^2\rho u)=r^2\dot q,
\end{equation}
\begin{equation}
\label{new.momentum}
u\frac{{\rm d}u}{{\rm d}r}+\frac{c^2}{\rho}\frac{{\rm d}\rho}{{\rm
d}r}=-\frac{\dot{q}u}{\rho},
\end{equation} 
where $u$ and $\rho$ are respectively the flow density and velocity.  
With the definitions
\begin{equation}
M=\frac{u}{c}\/,~{\tilde r}=\frac{r}{\Rf} \quad \mbox{and} \quad
{\tilde \rho}=\rho\frac{c}{\dot{q_0}\Rf} ,
\end{equation}
equations~(\ref{new.mass}) and~(\ref{new.momentum}) give
\begin{equation}
\label{dmdr}
\frac{{\rm d}M}{{\rm d}\tilde r}=\frac{M}{(M^2-1)}\left\{\frac{2}{\tilde r}-\frac{{\tilde r}^{\alpha}M^{\beta-1}(M^2+1)}{\tilde \rho}\right\},
\end{equation}
\begin{equation}
\label{drhodr}
\frac{{\rm d}\tilde\rho}{{\rm d}\tilde r}=\frac{2M^2}{(M^2-1)}\left\{ {\tilde r}^{\alpha}M^{\beta-1}-\frac{\tilde\rho}{\tilde r}\right\}.
\end{equation}
\subsection{Mach number independent mass injection}
As noted above, this case is appropriate to the injection of
material by photoionization, so that $\beta=0$ and
equations~(\ref{dmdr}) and~(\ref{drhodr}) give
\begin{equation}
M{\tilde\rho}{\tilde r^2}=\frac{\tilde r^{\alpha+3}}{(3+\alpha)},
\end{equation}
\begin{equation}
\label{diffeq}
\frac{{\rm d}M}{{\rm d}\tilde r}=\frac{M}{{\tilde r}(1-M^2)}\left\{(1+\alpha)+(3+\alpha)M^2\right\}.
\end{equation}
For $\alpha\neq -1$, the velocity and density distributions are
\begin{equation}
\label{alphar}
A{\tilde r}=\frac{M^{1/(1+\alpha)}}{\left| (1+\alpha)+(3+\alpha)M^2
\right| ^\gamma},
\end{equation}
\begin{equation}
\label{alpharho}
\tilde\rho=\frac{1}{A^{1+\alpha} \left|(1+\alpha)+(3+\alpha)M^2 \right|^\delta (3+\alpha)},
\end{equation}
where
\begin{displaymath}
\gamma=\frac{2+\alpha}{(3+\alpha)(1+\alpha)}\quad \mbox{and} \quad\delta=\frac{2+\alpha}{3+\alpha}.
\end{displaymath}
In the case $\alpha=-1$, equations~(\ref{alphar})
and~(\ref{alpharho}) become
\begin{equation}
A{\tilde r}=M^{-1/2}\exp(-1/(4M^2)) \quad \mbox{and} \quad
\tilde\rho=1/{2M}.
\end{equation}
$A$ is determined by the Mach number at the recombination front,
i.e. by $M=M_{\rm R}$.  If $M_{\rm R}<1$ then the \uchiir\ is pressure
confined by the external medium.
If the external pressure is small, which we adopt as the most probable
 case, then the flow becomes transonic at the recombination front
 (cf.\ Williams \& Dyson 1994) so that $A=(4+2\alpha)^{-\gamma}$
 ($\alpha\neq1$) and $A=\exp(-1/4)$ ($\alpha=-1$).
\begin{figure}
\centering
\mbox{\epsfxsize=8cm\epsfbox[0 400 535 735]{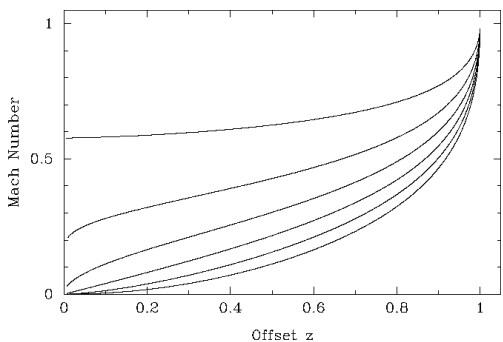}}
\caption{The variation of the Mach number with fractional offset $z\equiv r/\Rf$. From the lowest to uppermost curves, these correspond to $\alpha=1,~1/2,~0,~-1/2,~-1~$and$~-3/2.$ }
\label{rplot}
\end{figure}  
\begin{figure}
\centering
\mbox{\epsfxsize=8cm\epsfbox[0 400 535 735]{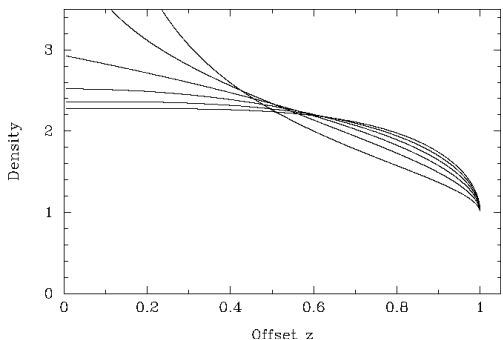}}
\caption{The variation of $\rho$, in units of $c(3+\alpha)/\dot{q_0}\Rf$, with $z$. From the lowest to uppermost curves (looking at the left hand side of the figure), these correspond to $\alpha=1,~1/2,~0,~-1/2,~-1~$and$~-3/2.$ }
\label{rhoplot}
\end{figure}

The Mach number and density distributions are given in
Figs~\ref{rplot} and~\ref{rhoplot} respectively, as a function of the
fractional offset $z=r/\Rf$, for six different values of $\alpha$
(including $\alpha=0$, which is uniform mass loading).  The densities
are normalized such that the edge density is unity for each $\alpha$
(i.e. so $\tilde\rho=1/(3+\alpha)$ at $M=\tilde{r}=1$).

\begin{figure}
\centering
\mbox{\epsfxsize=8cm\epsfbox[0 400 535 735]{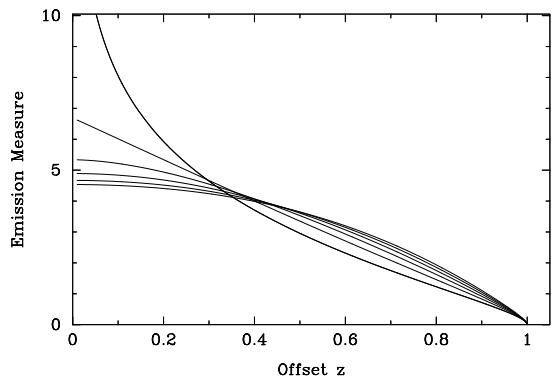}}
\caption{Emission measures as a function of fractional offset, $z$. The $\alpha=-3/2$ plot is the most centre-brightened, while the $\alpha=1$ plot is the least centre brightened.}
\label{remission}
\end{figure}

The emission measure, ${\rm EM}=\int n^2{\rm d}l$, is given in
Fig.~\ref{remission} as a function of the fractional offset, $z$, from
the central star, for each $\alpha$.  The individual plots are
calculated using the density distributions of Fig.~\ref{rhoplot}.  The
$\alpha=-3/2$~and~$-1$ distributions are centre-brightened and
reproduce a core--halo morphology.  A photoevaporating disc, as
described by Hollenbach et al.\ \shortcite{hollenbach.et.al94}, could
provide additional mass injection close to the central star.  However,
on its own, a photoevaporating disc gives an approximately $r^{-2}$
fall-off away from the disc, so additional mass injecting sources
would be needed away from the centre.  The other plots show
progressively a more uniform brightness with increasing $\alpha$.
These other models could reproduce satisfactorily \uchiir\ that have
spherical uniform morphologies.
  
\begin{figure}
\centering
\mbox{\epsfxsize=8cm\epsfbox[0 400 535 735]{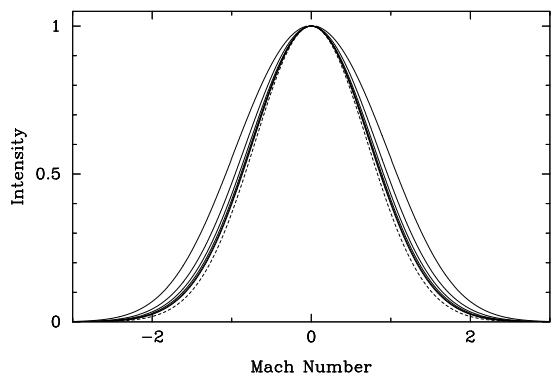}}
\caption{The profiles are almost identical but broaden slightly with decreasing $\alpha$ so that the outermost profile corresponds to $\alpha=-3/2$. The dashed curve, shown for comparison, has no Doppler broadening. The fractional offset $z=0.1$.}
\label{rprofile}
\end{figure}

The corresponding optically thin recombination line profiles are shown
in Fig.~\ref{rprofile}.  We assume that the gas emits with a Gaussian
profile, and we neglect effects such as pressure broadening
\cite{roelfsema&goss92} which will add characteristic Lorentzian wings
to the profiles.  They are viewed through lines of sight with offsets
of $z=0.1$ from the central star and are individually normalized to
their peak intensity.  The $\alpha=-3/2$~is distinct but is only
slightly broader than the higher $\alpha$ plots which are practically
identical and appear almost superimposed in the figure.  This shows
that, in this case, line profiles will not be able to differentiate
between the models described above.  These Gaussian-like profiles are
produced simply because the gas everywhere has a subsonic velocity.
%
\subsection{Mach number dependent mass injection}
\begin{figure}
\centering
\mbox{\epsfxsize=8cm\epsfbox[0 400 535 735]{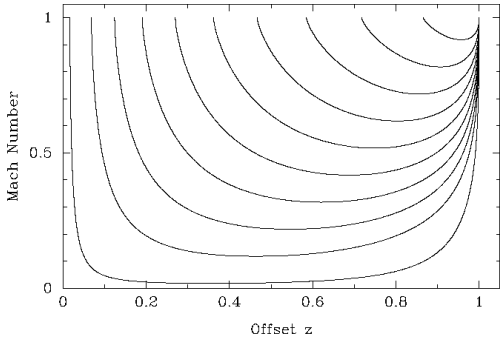}}
\caption{Family of Mach number solutions plotted as a function of offset $z$ for $\alpha=0$. We set $M=1$ at the recombination front at $z=1$ as one boundary condition and vary the inner boundary condition.}
\label{mach_family}
\end{figure}

\begin{figure}
\centering
\mbox{\epsfxsize=8cm\epsfbox[0 400 535 735]{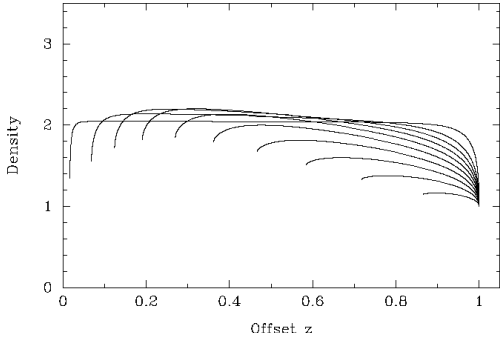}}
\caption{The corresponding density solutions plotted as a function of $z$ for $\alpha=0$. The initial conditions are the same as in Figure 5.}
\label{rho_family}
\end{figure}

\begin{figure}
\centering
\mbox{\epsfxsize=8cm\epsfbox[0 400 535 735]{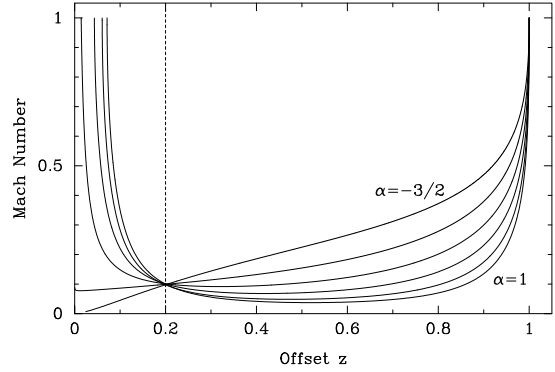}}
\caption{The Mach number plotted as a function of offset $z$. The two extreme values of $\alpha$ are indicated. The initial Mach number is $M=0.1$ at $z=0.2$ and we also set $M=1$ at $z=1$.}
\label{mach_ablated}
\end{figure}

\begin{figure}
\centering
\mbox{\epsfxsize=8cm\epsfbox[0 400 535 735]{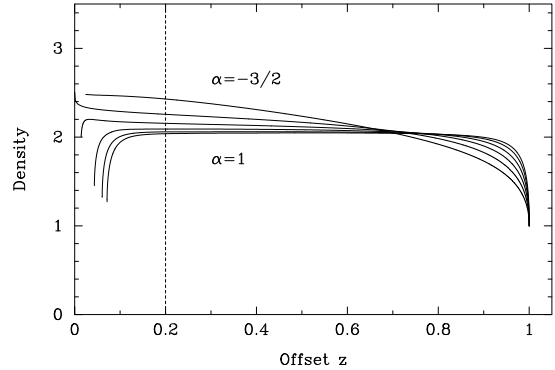}}
\caption{The density plotted as a function of $z$. The initial conditions are the same as in Figure 7.}
\label{rho_ablated}
\end{figure}

We now take the mass injection rate to be proportional to $M^{4/3}$
which is appropriate for mass loading by hydrodynamical ablation.
Substituting $\beta=4/3$ into equations~(\ref{dmdr})
and~(\ref{drhodr}) gives
\begin{equation}
\frac{{\rm d}{M}}{{\rm d}\tilde{r}}=\left ( \frac{M}{M^2-1}\right ) \left [\frac{2}{\tilde{r}}-\frac{M^{1/3}(1+M^2)\tilde{r}^\alpha }{\tilde\rho}\right ],
\end{equation} 
\begin{equation}
\frac{{\rm d}{\tilde\rho}}{{\rm d}\tilde{r}}=\frac{2M^2}{(M^2-1)}\left [ M^{1/3}\tilde{r}^\alpha -\frac{\tilde{\rho}}{\tilde{r}}\right ].
\end{equation}
There is no non-trivial solution for hydrodynamic ablation that passes
through $\tilde{r}=0$, so we assume that the flow is started off by
another mechanism such as photoionization, or by a stellar wind
shocking, and match this at some radius to our model here.  As one of
our boundary conditions we have $M=1$ at $\tilde{r}=1$.  Varying the
other boundary condition (e.g. $\tilde{\rho}$ at $\tilde{r}$) generates
a family of solutions for the velocity and density for each $\alpha$.
We present these solutions in Figs~\ref{mach_family}
and~\ref{rho_family} for the case of $\alpha=0$ only, for clarity.
To show the variation of these solutions with $\alpha$, we select the
solution that passes through the point at $\tilde{r}=0.2$ with $M=0.1$
as an example.  We retain the other boundary condition $M=1$ at
$\tilde{r}=1$ and examine the same six power laws as before.  The full
velocity and density solutions for this particular pair of boundary
conditions are shown in Figs~\ref{mach_ablated} and~\ref{rho_ablated}.
The densities are normalized to the same edge density as before.  The
variation of the flows between different values of $\alpha$ is less
marked than in the Mach number independent models of the previous
subsection.  The Mach number dependence of the mass injection rate
means that mass loading is enhanced in a shell towards the edge of the
region.  This effect is most pronounced for the positive values of
$\alpha$.  Gravitational effects in fact may become important at the
parts of the flow with particularly low velocities or at small radii,
and we will discuss this elsewhere.

\begin{figure}
\centering
\mbox{\epsfxsize=8cm\epsfbox[0 400 535 735]{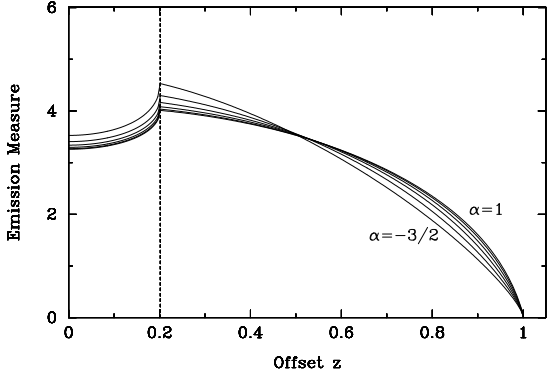}}
\caption{Emission measures for the same initial conditions and radial dependences as above. The radius at which the flow begins is indicated by the dashed line.}
\label{emissions_alpha}
\end{figure}

\begin{figure}
\centering
\mbox{\epsfxsize=8cm\epsfbox[0 400 535 735]{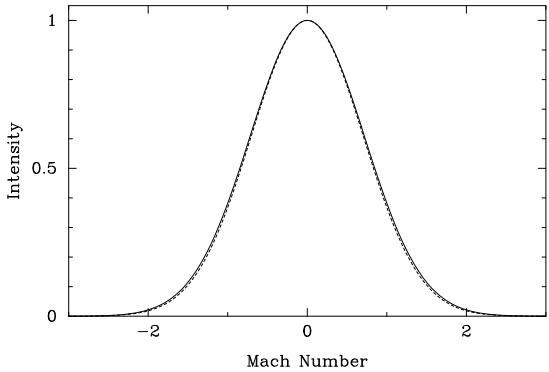}}
\caption{An example of a line profile for the same initial conditions and radial dependences as above with $z=0.2$. The six profiles are practically identical so we show just one here. The dashed curve, shown for comparison, has no Doppler broadening.}
\label{profiles_alpha}
\end{figure}

In Fig.~\ref{emissions_alpha} we show the emission measure as a
function of fractional offset $z$.  Since there may be several ways in
which the flow is initiated, we neglect any emission from this inner
zone which is not part of the solution.  This gives the dip for
offsets less than $z=0.2$, since our line of sight contains a sphere
with no emission.  It can be seen that the emission measures are quite
similar for the different radial fall-offs.  Comparing these with the
emission measures for the flow loaded by photoionization ablation
(shown in Fig.~\ref{remission}), we see that the Mach number dependent
models are similar to the photoionization loaded flows with $\alpha\ga
0$, and thus they could also describe a \uchiir\ with uniform
morphology.  An optically thin line profile is shown in
Fig.~\ref{profiles_alpha} for a fractional offset $z=0.2$.  The
profile is hard to distinguish from the profiles of the
photoionization loaded flows with $\alpha\ga 0$ seen in
Fig.~\ref{rprofile}.  Only one profile is shown because the six models
give practically the same line profiles.
%

\section{\uchiir\ in terms of a supersonic--subsonic flow with an internal shock}
In this section, we combine the results of the previous sections with
those of Paper I to allow for the possibility that the flow is
decelerated by mass loading to such an extent that a wind termination
shock occurs before the recombination front is reached.  This will
occur if the Mach number in the supersonic flow is predicted by
the ballistic approximation to reach values of $M\la 2$ before the
recombination front \cite{williams.et.al95}.  In Paper I we used the
conservation of mass and momentum to find velocity and density
distributions for a fully supersonic, uniformly mass loaded flow:
\begin{equation}
\label{super}
M=\frac{3\dot{\mu_*}}{4\pi\dot{q}r^3c};\quad
\rho=\frac{4\pi\dot{q}^2r^4}{9\dot{\mu_*}}.
\end{equation}
The mass loading mechanism in the supersonic flow before the shock
radius is independent of Mach number \cite{hartquist.et.al86}.  The
emission will be dominated by the low-velocity, high-density gas at
the edge of the mass loading region.  The high velocity of the
unshocked wind will mean that it stands well clear of the
near-Gaussian profiles predicted for the subsonic region.  However in
all but the cases with the lowest pre-shock velocities, or thinnest
subsonic shells, the high density in the subsonic region (which is a
factor substantially more than 4 higher than the pre-shock density
because of the cooling of the shocked gas) will mean that this
low-velocity gas dominates the emission.  For simplicity, we assume
that the mass injection rate is independent of flow velocity and
radial distance so that $\beta=\alpha=0$.

\begin{figure}
\centering
\mbox{\epsfxsize=8cm\epsfbox[0 400 535 735]{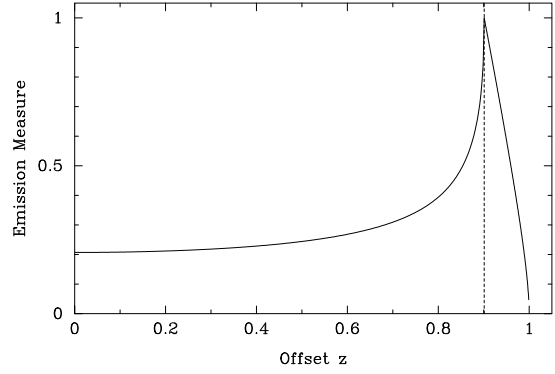}}
\caption{Emission measure as a function of $z$ for a supersonic--subsonic flow. The position of the shock is indicated by the dashed line.}
\label{fatemission}
\end{figure}

\begin{figure}
\centering
\mbox{\epsfxsize=8cm\epsfbox[0 400 535 735]{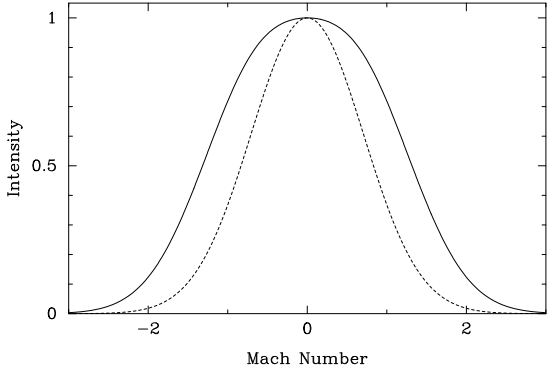}}
\caption{Line profile for $z=0.1$ taken through a region with a shock at $0.9\Rf $. The dashed curve, shown for comparison, has no Doppler broadening.}
\label{fatprofile}
\end{figure}

In the emission measure plot shown in Fig.~\ref{fatemission} a shell
morphology is clearly apparent.  The emission at offsets less than the
shock radius is from the subsonic gas in our line of sight and not
from the supersonic gas which contributes very little to the emission.
In fact, even if there were noticeable emission from the supersonic
gas, the overall emission measure would still be essentially that of a
shell because the density distribution (equation~\ref{super}) is
strongly weighted to the edge of the supersonic region.  In
Fig.~\ref{fatprofile} we show an optically thin line profile for this
shock radius taken at a line of sight with offset $z=0.1$, close to
the centre.  The profile consists of two near-Gaussian components,
generated by the parts of the shell that are moving towards and away
from us, separated in velocity by an amount that is a fraction of 
their widths.  This results in a line profile that is broader than
those produced using the fully subsonic models described earlier.  We
find very similar results if the mass loading is Mach
number dependent.

\section{conclusions}
In this paper we have modelled \uchiir\ as subsonic flows from
photoionized clumpy clouds lying in the vicinity of an ionizing
stellar source.  We have found that, by assuming a simple spatial
distribution of clumps and allowing the flow to be supersonic,
subsonic or both (with a separating shock), we can reproduce three
different \uchiir\ morphologies.

For the fully subsonic models, we have found that a radial dependence
of the mass injection rate is more important for the
photoionization-induced subsonic flow.  The emission measures show
that the $M^{4/3}$ flows are less centre-brightened than the
photoionization flow.  Either of the two types of model can naturally
describe the uniform spherical \uchiir.  In addition, the
photoionization flow with $\alpha=-3/2$~or $-1$ can satisfactorily
explain the core--halo \uchiir.  Line profiles are of no use in
distinguishing between the two different flows.  For both cases we
have calculated velocity and density distributions that will be used
in more detailed future work.

Paper I showed that a supersonic wind-blown model could reproduce the
shell morphology but with broad, double-peaked line profiles.  In
contrast, we have suggested in Section 3 that if a shock forms before
the recombination front then a shell morphology is retained but with a
single peaked line profile.  Which type an individual object is will
be settled by high-resolution observations of the velocity structure
of the region.

At high enough resolution, the largest individual clumps will begin to
be resolved.  The irregular/multiple-peaked \uchiir\ are perhaps
simply a collection of large clumps which are ionized on their
outsides by the central source.  The `hypercompact' continuum sources
found in clusters by Gaume et al.\ \shortcite{gaume.et.al95} are
perhaps clumps such as this.  Preliminary work suggests that a
non-spherically symmetric distribution in the mass loading around the
region can describe the arc-like \uchiir\ (Paper III).

The solutions to the subsonic flow equations may be applicable to
other objects.  Although visible \hii\ regions are older and more
evolved than \uchiir, their morphologies have been found to be very
similar to those of \uchiir\ \cite{fich93}, and this may imply that
they could be used to test some of the observational diagnostics
described here and in Paper I\@.  The extent of the neutral H\,{\sc i}
envelopes around \hii\ regions is explained by the structure of
photodissociation regions in clumpy clouds
\cite{howe.et.al91}. However, Roger \& Dewdney
\shortcite{roger&dewdney92} note that the outflowing H\,{\sc i} shocks
seen around \hii\ regions (e.g. Kuchar \& Bania 1993) are not as thin
and do not have as high densities as predicted by standard theory.
Such low-density shells of H\,{\sc i} may result from support by the
swept-up magnetic fields \cite{mathews&odell69}. We would suggest that
the neutral winds beyond recombination fronts would also explain these
structures. If this is the case, then observations of \hii\ regions
agreeing with our predictions for \uchiir\ would lend support to our
models.

We will address elsewhere the dynamical effects that dust will
have on these models, consider the intermediate-scale structure in the
regions and incorporate any gravitational effects.
 
\section*{Acknowledgments}

This work was supported by PPARC both through the Rolling Grant to the
Astronomy Group at Manchester (RJRW) and through a Graduate
Studentship (MPR).

\label{lastpage}
\end{document}